\documentclass[a4paper,referee]{raa}
\DeclareUnicodeCharacter{FF0C}{,}
\usepackage{graphicx,times}      
\usepackage{natbib}
\usepackage{amssymb,amsmath}
\usepackage{soul}
\usepackage{xcolor}
\bibpunct{(}{)}{;}{a}{}{,}
\usepackage[a4paper=true,pagebackref=true]{hyperref}
\hypersetup{colorlinks = true, linkcolor = green, anchorcolor = red, citecolor = blue, filecolor = red, pagecolor = red, urlcolor = red}

\DeclareGraphicsExtensions{.ps,.eps.gz,.epsi}
\DeclareGraphicsExtensions{.ps,.eps.gz,.epsi}

\hypersetup{
    colorlinks=true,
    linkcolor=blue,
    filecolor=magenta,      
    urlcolor=blue,
    pdftitle={Overleaf Example},
    citecolor=green,
    pdfpagemode=FullScreen,
    }

\begin{document}

   \title{The Mini-SiTian Array: the mini-SiTian Realtime Image Processing pipeline (STRIP)}

   \volnopage{Vol.0 (20xx) No.0, 000--000}      
   \setcounter{page}{1}          

   \author{\href{https://orcid.org/0009-0007-5610-6495}{Hongrui Gu}
      \inst{1,2}
   \and \href{https://orcid.org/0000-0003-3250-2876}{Yang Huang}$^{\ast}$
      \inst{1,2}
   \and \href{https://orcid.org/0000-0002-3935-2666}{Yongkang Sun}
      \inst{1,2}
    \and \href{https://orcid.org/0000-0001-8424-1079}{Kai Xiao}
      \inst{1,2}
   \and \href{https://orcid.org/0000-0002-7598-9250}{Zhirui Li}
      \inst{1,2}
    \and \href{https://orcid.org/0009-0001-5324-2631}{Beichuan Wang}
      \inst{1,2}
    \and \href{https://orcid.org/0000-0002-6790-2397}{Zhou Fan}$^{\ast}$
      \inst{1,2}
   \and \href{https://orcid.org/0009-0006-7556-8401}{Chuanjie Zheng}
      \inst{1,2}
   \and \href{https://orcid.org/0000-0003-3474-5118}{Henggeng Han}
      \inst{1,2}
    \and \href{https://orcid.org/0000-0002-6684-3997}{Hu Zou}
      \inst{1,2}
    \and \href{https://orcid.org/0000-0002-0096-3523}{Wenxiong Li}
      \inst{1,2}
    \and \href{https://orcid.org/0000-0002-4333-3994}{Hong Wu}
      \inst{1,2}
    \and \href{https://orcid.org/0000-0002-2874-2706}{Jifeng Liu}
      \inst{1,2}
   }

   \institute{CAS Key Laboratory of Optical Astronomy, National Astronomical Observatories, Chinese Academy of Sciences, Beijing 100101, People's Republic of China; \\
    \and
    School of Astronomy and Space Science, University of Chinese Academy of Sciences, Beij\textbf{}ing 100049, People's Republic of China\\
    Corresponding author: Yang Huang (huangyang@ucas.ac.cn); Zhou Fan (zfan@nao.cas.cn)\\
\vs\no
   {\small Received~~2024.11.11; accepted~~2025.3.17}}

\abstract{
This paper provides a comprehensive introduction to the Mini-SiTian Real-Time Image Processing pipeline (STRIP) and evaluates its operational performance. The STRIP pipeline is specifically designed for real-time alert triggering and light curve generation for transient sources. By applying the STRIP pipeline to both simulated and real observational data of the Mini-SiTian survey, it successfully identified various types of variable sources, including stellar flares, supernovae, variable stars, and asteroids, while meeting requirements of reduction speed within 5 minutes. For the real observational dataset, the pipeline detected 1 flare event, 127 variable stars, and 14 asteroids from three monitored sky regions. Additionally, two datasets were generated: one, a real-bogus training dataset comprising 218,818 training samples, and the other, a variable star light curve dataset with 421 instances. These datasets will be used to train machine learning algorithms, which are planned for future integration into STRIP.
}

   \authorrunning{Gu et al.}            
   \titlerunning{SiTian Pioneer Project–Mini-SiTian IV: STRIP}  

   \maketitle

%
%
\section{Introduction}           
\label{sect:intro}

Transient sources are often associated with extreme physical processes. Investigating their early evolution can validate or expand our astrophysical theories under extreme astronomical phenomena, such as solar flares (\citealt{2015ApJ...809..104A}), kilonovae, and supernovae (\citealt{2024Natur.627..754L}). To study these transient events, time-domain surveys like the Lick Observatory Supernova Search (LOSS, \citealt{2000AIPC..522..103L}, \citealt{2001ASPC..246..121F}), began observing the sky towards the end of the last century. Subsequently, the emergence of projects such as the Palomar Transient Factory (PTF, \citealt{2009PASP..121.1395L}, the predecessor to the currently operational Zwicky Transient Facility, ZTF, \citealt{2019PASP..131a8002B}), the Catalina Real-Time Transient Survey (CRTS, \citealt{2009ApJ...696..870D}, \citealt{2011arXiv1102.5004D}), the Asteroid Terrestrial-impact Last Alert System (ATLAS, \citealt{2011tfa..confE..24T}, \citealt{2018PASP..130f4505T}), the All-Sky Automated Survey for Supernovae (ASAS-SN, \citealt{2014AAS...22323603S}, \citealt{2017PASP..129j4502K}),  and other projects significantly broadened human understanding of various types of transient sources.

However, single-band time-domain surveys provide limited information. Since transient events are often non-repeating and short-lived, missing the early burst means that crucial temperature and spectral line features cannot be adequately captured using only single-band data. To address this limitation, numerous simultaneous multicolor observation programs and high-cadence time-domain surveys have been developed to obtain more comprehensive information on transients. Examples include the Multi-channel Photometric Survey Telescope (Mephisto, \citealt{2019gage.confE..14L}), the Tsinghua University-Ma Huateng Telescopes for Survey (TMTS, \citealt{2020PASP..132l5001Z}), and the Ground Wide Angle Cameras Array (GWAC-A, \citealt{2021PASP..133f5001H}), which achieve minute-level sampling. However, existing wide-field, high-cadence telescope arrays generally have shallow detection limits, while large-aperture multicolor telescopes often lack sufficient sampling frequency. To overcome these challenges, the SiTian project (\citealt{2021AnABC..93..628L}) was initiated. Consisting of dozens of 1-meter-aperture telescopes, each with a 25-square-degree field of view, SiTian is designed to capture simultaneous three-color images at a depth of magnitude 21 every 30 minutes, significantly enhancing our ability to study rapid and faint transient sources.

Nevertheless, time-domain surveys using only a few broad-band filters provide far less information than spectroscopy. To fast capture the nature of transient sources, near-real-time processing of time-domain photometric images is essential, enabling a rapid response for multi-band follow-up observations, including optical spectroscopy, radio, and high-energy observations.

In optical imaging, high-precision aperture photometry combined with catalog cross-matching can quickly and effectively identify variations of variable sources in non-crowded fields, thereby triggering follow-up spectroscopic observations or even observations in other wavelengths, such as radio and X-rays. However, many major transient sources, such as supernovae or tidal disruption events (hereafter TDEs), occur in crowded star fields or galactic disks, where aperture photometric measurements cannot be used.
To address this issue, algorithms like ISIS (\citealt{1998ApJ...503..325A}, \citealt{2000A&AS..144..363A}) , Hotpants (\citealt{2015ascl.soft04004B}), and SFFT (\citealt{2022ApJ...936..157H}) have been developed for image subtraction. Point-like features remaining in the post-subtraction image indicate variable sources, while the invariant, crowded stellar fields or galactic disks are removed, thus enabling the discovery and measurement of variable sources through aperture photometry.

The Mini-SiTian array serves as the pathfinder for the SiTian project, utilizing the same observational mode to function as an exploration and pre-research platform. To facilitate rapid responses to transients during observations, it is essential to have a dedicated transient response pipeline based on image subtraction techniques. This study, which is based on the Hotpants algorithm, has been tested on Mini-SiTian Field-02 (F02) and selected supernova fields. The tests have verified the pipeline's feasibility and robustness, successfully identifying a range of variable sources. The primary focus of this paper is to introduce the complete pipeline and demonstrate its effectiveness through the results obtained from these tests.

The structure of this paper is as follows: Section 2 provides an overview of the workflow for the Mini-SiTian real-time image processing pipeline (STRIP). Section 3 evaluates the performance of the pipeline using both simulated and real observational data. Section 4 presents the data products generated from the test observations. Finally, Section 5 summarizes the key findings, discusses the accomplishments of the pipeline, and outlines the anticipated future directions for Mini-SiTian operations.

\section{STRIP pipeline}
\label{sect:Data}
The Mini-SiTian telescopes are currently operational at the Xinglong Observatory. It consists of three identical 300mm f/3 telescopes, equipped with Sloan-like g', r', and i' band filters, and designed to simulate a single node of the SiTian array. The terminal instrument used is an ASI6200 CMOS camera from ASI, which maintains the same basic CMOS structure as the cameras in the SiTian array. This configuration provides a field of view (FoV) of 2.29° $\times$ 1.53°, with a pixel resolution of 9576 $\times$ 6388, corresponding to 0.862 arcseconds per pixel. For more detailed information about the telescope, refer to \cite{10.1088/1674-4527/adc788}, and for further details about the CMOS cameras, refer to Zhang et al. (2025). 

During operations, the full width at half maximum (FWHM) of images typically ranges between $2''$ and $4''$. For all fields except those designated for gravitational wave events,  the Mini-SiTian telescopes operate in staring mode, consistently targeting the same patch of the sky. Differential images are generated by subtracting these observations from a high signal-to-noise ratio (SNR) reference image, which is compiled by stacking quality-filtered historical observations of the same region. This approach allows for the identification of sources that exhibit potential photometric variations. 

When a source is identified as real--i.e., not an artifact caused by cosmic rays or instrumental errors--based on specific criteria or algorithms, and it is not listed in any known variable star database, it is classified as a potential transient event. An alert is then triggered, prompting follow-up observations with larger telescopes to further investigate the phenomenon.

For transient search tasks within a fixed sky area, such as detecting electromagnetic counterparts of gravitational wave events where prior information about the sky region is available, the Mini-SiTian array will adopt a scanning mode. In this mode, the array observes segment of sky for a few exposures before moving on to the next segment. Due to the large combined field of view (FOV) of the three-color nodes in the SiTian project, it is possible to generate reference images of the entire sky relatively quickly. Consequently, the method used for transient searches in scanning mode is similar to that in staring mode. 

However For the Mini-Sitain project, which represents only a single node of SiTian array, the FoV of a single telescope is much smaller (approximately 3 deg$^2$ for a Mini-SiTian telescope, compared to about 25 deg$^2$ for a SiTian telescope). Generating a reference images for the entire sky is not feasible during the test phase and would require roughly one year to complete. As a result, this work focuses on evaluating the performance of the staring mode rather than testing transient search tasks in scanning mode.

This section presents a flowchart outlining the STRIP pipeline for transient detection (see Figure 1) and discusses the rationale behind the design choices made for this pipeline.

\subsection{Image Processing}
Prior to daily observations, calibration frames, including flats and biases, are obtained. Preprocessing begins by stacking the bias and flat-field images from the day, resulting in the creation of a master bias and a master flat for calibration purposes. This step ensures uniform background levels and consistent pixel response across images. Without it, polynomial fitting of the background during the subsequent HOTPANTS image subtraction process can lead to uneven background levels, which may negatively affect star detection by SExtractor.

Once generated, these master frames are archived in a dedicated repository. In cases where adverse weather conditions or other factors prevent the capture of flat and bias frames on a given day, the pipeline retrieves previously archived master calibration frames to replace the missing ones. Since the characteristics of flat-field and bias frames for the CMOS cameras used in the Mini-SiTian array do not change rapidly over time (Zhang et al., 2025), recent calibration frames are generally sufficient for rapid processing and transient source detection. Consequently, the pipeline automatically retrieves the most chronologically recent master calibration frames from the repository for calibration.

Upon completion of observations, the images are transmitted to the server hosting the STRIP pipeline, which first verifies the integrity of the received data. Following successful transmission, the STRIP pipeline automatically performs preprocessing steps, including bias-field and flat-field corrections. After preprocessing, the source extraction process is carried out using SExtractor (\citealt{1996A&AS..117..393B}), with sources detected at a signal-to-noise ratio exceeding two sigma above the background. Images containing more than 100 stars detected by SExtractor proceed to have their World Coordinate System (WCS) information solved using \href{astrometry.net}{`astrometry.net'} (\citealt{2010AJ....139.1782L}). In cases of poor weather conditions or sparse star fields, where fewer than 100 stars are detected, the WCS solution may fail. In such instances, the affected images are excluded from further processing.

The calibrated images with updated WCS information possess a certain level of astrometric precision. However, the error associated with this astrometric solution typically remains several pixels, which does not meet the precision requirements for accurate image subtraction. Therefore, further refinement of the astrometric solution is necessary before storing the images in good image repository. To achieve this, we first use the derived WCS information to extract bright stars (brighter than 18.5 magnitudes) from the Gaia catalog in the nearby sky area, saving them in LDAC format. We then employ SCAMP (\citealt{2006ASPC..351..112B}) to perform a more precise WCS refinement using the LDAC catalog. This refined WCS is subsequently used by SWarp (\citealt{2010ascl.soft10068B}) to resample the image onto the coordinate grid defined by the template image. If this image is the first to be obtained for a given target or field, its WCS serves as the template WCS for that field, and the image is stored as the initial template in the template library for alignment purposes. All subsequent images are aligned to this template WCS. 

After resampling and alignment, the images are archived in a repository designated for good images. Once stored, the pipeline uses SExtractor to extract star catalogs and compares them with the Gaia DR3 catalog to calculate the magnitude zero point and the overall average FWHM. Additionally, the limiting magnitudes are estimated. This information is used to identify images with smaller seeing and higher limiting magnitudes, which are then selected for stacking to produce a template image with improved SNR.

The HOTPANTS image subtraction method involves convolving the template image with a series of kernels to match its point spread function (PSF) as closely as possible to that of the science image prior to subtraction. This convolution process generally increases the FWHM of the stellar profiles. To achieve a PSF in the template image that closely matches that of the science image, the FWHM of the template should be smaller than or equal to that of the science image. Therefore, images with smaller seeing are preferred for use as the template. 

Image subtraction is essentially a process in which the noise in the resulting image is the sum of the noise from both the template and the science images. To minimize noise in the subtracted image, it is crucial to obtain high SNR images by stacking multiple images with smaller seeing. Consequently, to achieve high SNR and minimized seeing, the template image should be updated regularly as new high-quality images become available in the image library. The template library will maintain updated through the survey proceeding. 

At the conclusion of each observation day, the pipeline automatically reviews the repository of high-quality images, selecting the image with an FWHM of less than 4.5 pixels (approximately 3.9 arcseconds). When more than 20 such images are available, the STRIP pipeline will select the top 75\% of these images, based on their limiting magnitude, for stacking. The resulting stacked image will then replace the old template and be stored as the updated template in the library. These selection criteria are adjusted according to the telescope's pixel size and the site’s seeing conditions to ensure that images with smaller seeing are selected, while maintaining a sufficient number of images for analysis.

This process encapsulates the comprehensive image preprocessing workflow. With preprocessed observational images and regularly updated templates, the staring mode can effectively identify variable stars within the observed fields, as further detailed in Section 2.2.

\subsection{Staring Mode Observations of Transient Sources}
The staring mode of the Mini-SiTian project is designed to detect fast-evolving transients, such as flares that last from minutes to hours, as well as the photometric variations of supernovae that evolve over days to months. To identify these transient sources, the STRIP pipeline utilizes image subtraction software within a CPU environment. After preprocessing and image alignment using SWarp, the images are prepared for comparison with the template image using the HOTPANTS software. By inputting both the template and the resampled science image into HOTPANTS, a residual image is generated, highlighting primarily variable sources. An example of the template image, science image, and residual image is shown in Figure 2.
\begin{figure*}
   \centering
    \includegraphics[width=15cm]{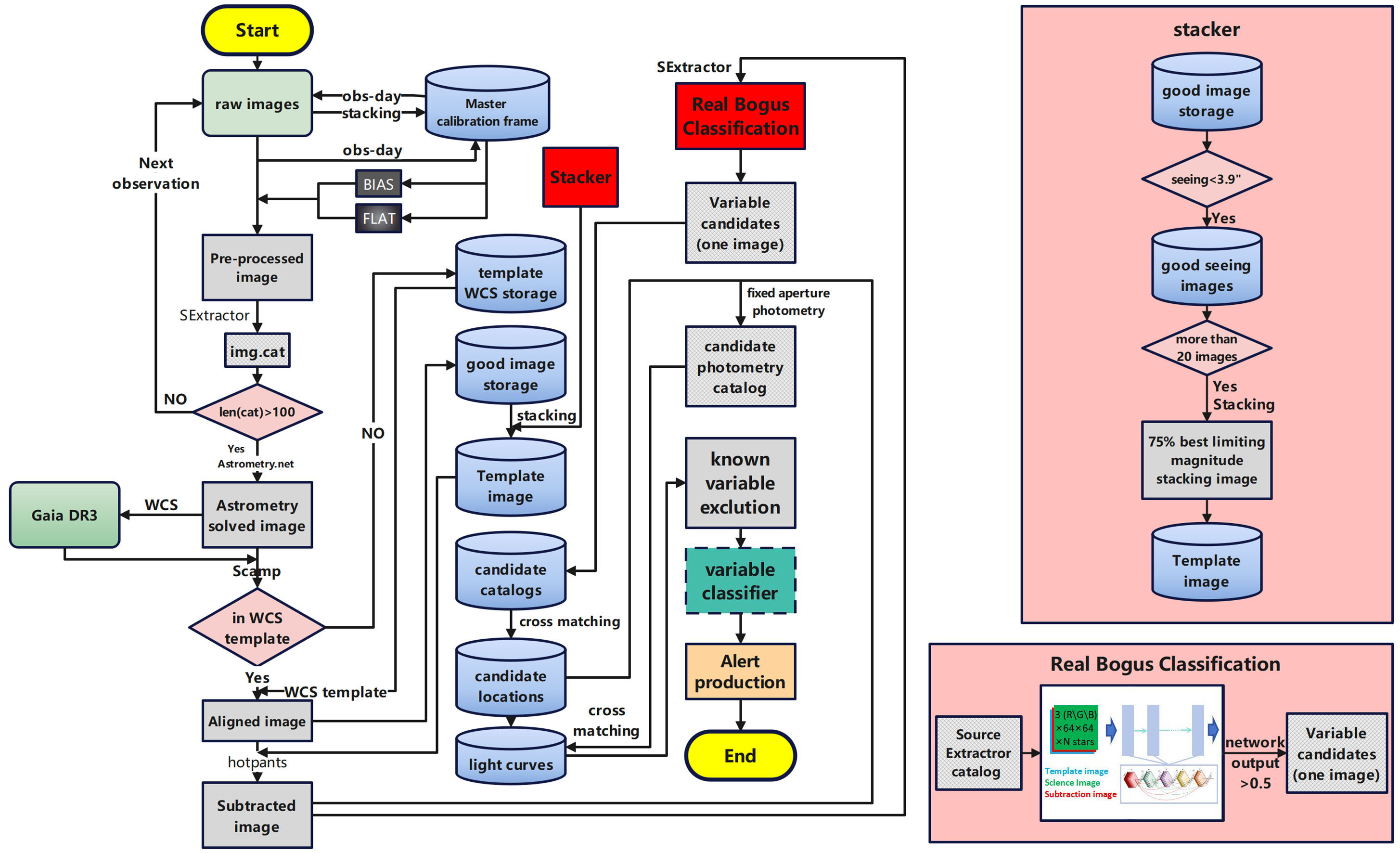}
   \caption{Flowchart of STRIP Pipeline (left panel), detailed flowchart of the stacker block (upper right), and detailed flowchart of the real bogus classifier (lower right).
   }
   \label{}
\end{figure*}
\begin{figure*}
   \centering
    \includegraphics[width=15cm]{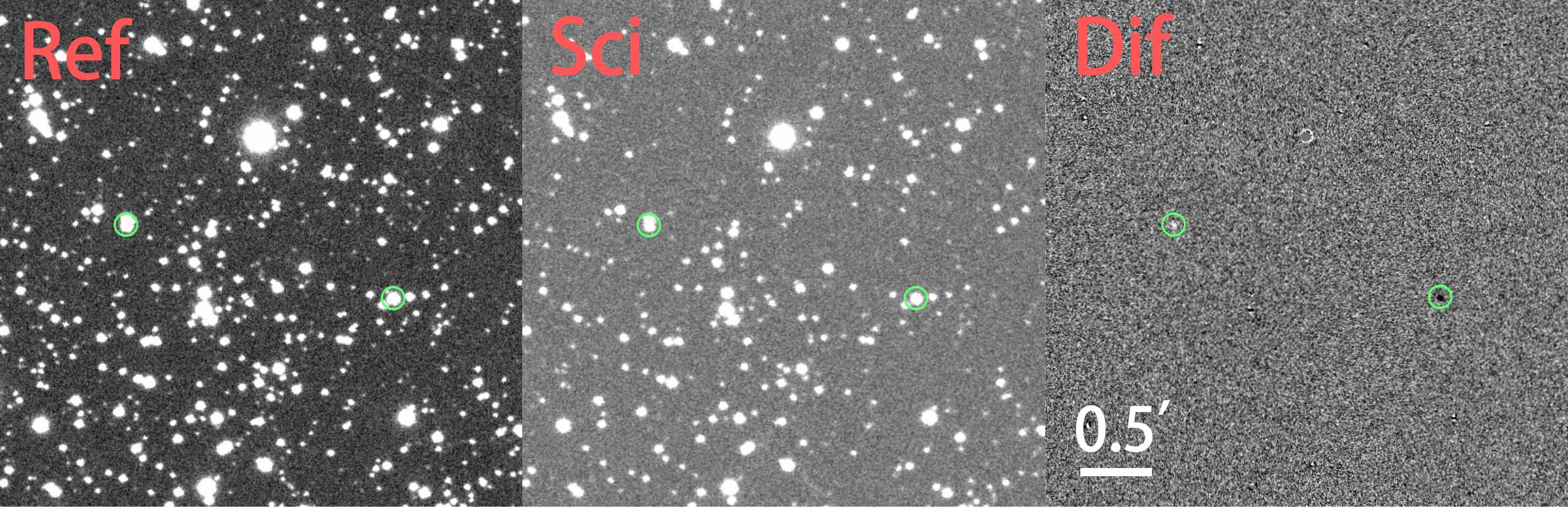}
   \caption{
   Comparison between reference image (left panel) , science image (middle panel) and subtraction image (right panel) in field-02 (F02) in Mini-SiTian test run. The subtraction image  is generated by subtracting reference image from science image.The variable sources in this sky region are highlighted in the green circles. The left star got brighter and the right one got dimmer.
   }
   \label{}
\end{figure*}
\begin{figure*}
   \centering
    \includegraphics[width=14.5cm]{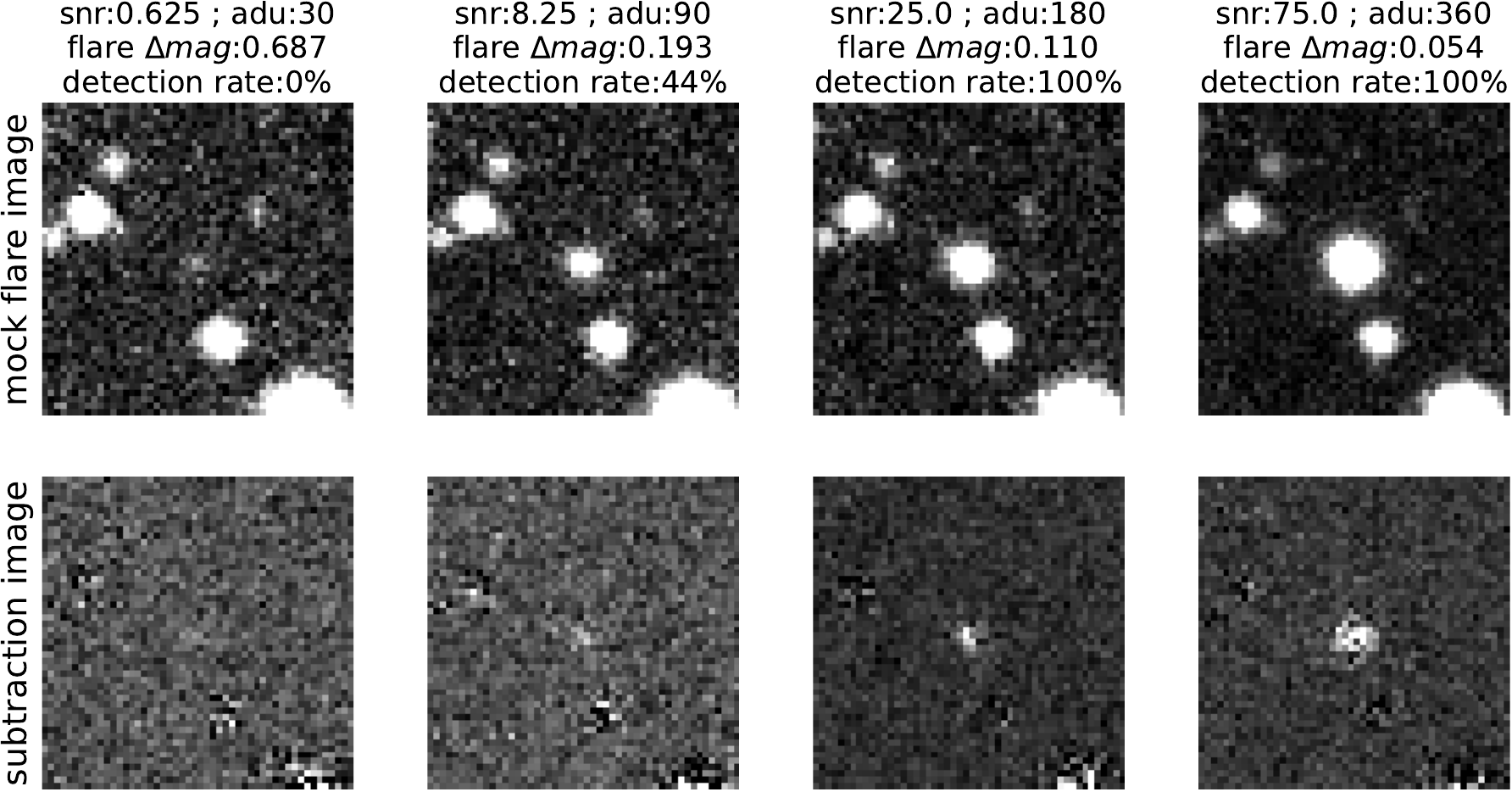}
   \caption{
   Performances of simulated stellar flares under different SNR and flare intensity. The four images above, from left to right, depict simulated observations of stellar flares for stars with SNRs of 0.625, 8.25, 25, and 75, respectively, producing flare peaks increasing 30, 90, 180, and 360 ADUs. The corresponding flare magnitude increments for these four events are 0.687, 0.193, 0.110, and 0.054 magnitudes. Under the processing by the STRIP pipeline, the detection rates for these four categories of flare events are 0\%, 44\%, 100\%, and 100\%. The four images below illustrate the subtraction images generated during the STRIP pipeline process. It is evident that the flares in the subtraction images become increasingly prominent as the peak ADU increases.
   }
   \label{}
\end{figure*}
\begin{figure*}
   \centering
    \includegraphics[width=14.5cm]{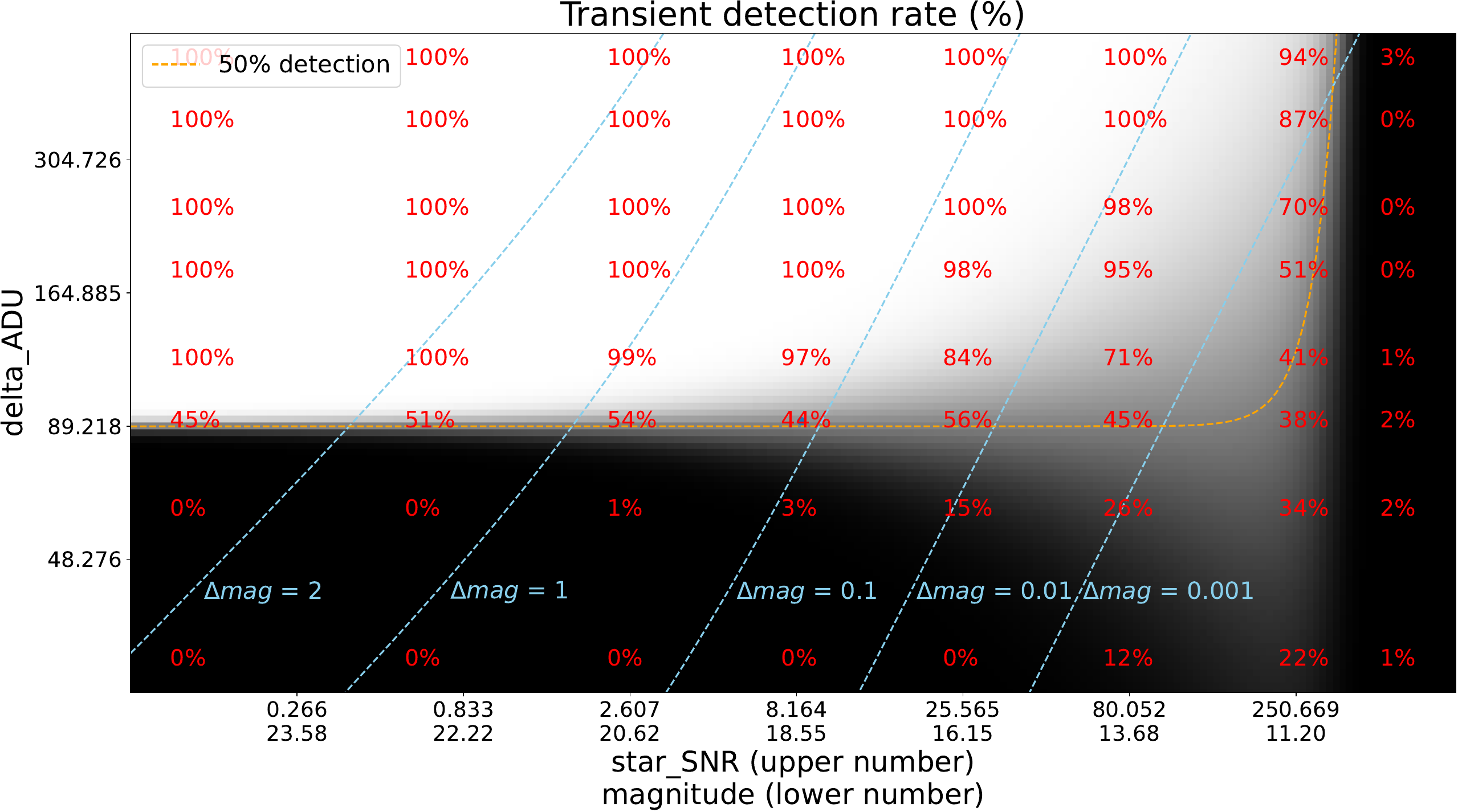}
   \caption{
   Detection rates of STRIP across host star with different brightness and flare intensities. The red numbers represent the detection rates of the STRIP pipeline corresponding to different combinations of SNRs (upper number of x-axis) and peak brightness increases in unit ADU (y-axis) for stellar flares. Each one examples in Figure 3 illustrate instances from one x-y combinations out of 100 simulations. For better understanding, the x-axis has been converted into magnitude values by comparing  with a 300-second exposure image, which 5 $\sigma$ limiting magnitude is 19.5 (lower number of x-axis). Additionally, the combination of star brightness and flare increments in unit ADU has been transformed into differential flare magnitude increments, which are indicated with blue dashed lines. The red values have been converted into a black and white detection rate distribution ranging from 0\% to 100\% by fitting the same x-data with a sigmoid function and then fitting the parameters of the sigmoid function with an exponential function. The orange curve represents the position where the detection rate is 50\%.
   }
\end{figure*}

Given the inherent instability of image subtraction methods and the potential for cosmic rays and saturated pixels to appear in the subtraction images, the STRIP pipeline employs a real-bogus machine learning classifier to refine the source extraction process. This approach, similar to that used in ZTF (\citealt{2019MNRAS.489.3582D}), helps mitigate these issues. The training and operation of the classifier are described in Section 4. Since SExtractor can only detect positive-valued sources, a second source extraction process is conducted after the residual map is inverted. The real-bogus classification is then repeated, and the intersection of positive and negative detections forms the candidate variable source catalog for the observation.

This candidate catalog is archived and cross-matched with previous candidate catalogs for the same field, using a specified radius of $3''$ to refine the list. The coordinates from this cross-match serve as the basis for fixed-aperture photometry on both the template and subtracted images, yielding magnitude measurements that are recorded in a light curve database. 

Promising candidates are then prioritized for follow-up observations with larger telescopes. The flexibility of the STRIP pipeline allows for the integration of alternative filters, such as cross-matching with galaxy catalogs to identify potential supernova candidates, highlighting the dynamic and adaptable nature of the pipeline in the context of transient astronomy.

\section{Operational Testing}
This section presents an empirical evaluation of the STRIP pipeline's performance in detecting two key transient phenomena—stellar flares and supernovae—during staring mode observations, utilizing both simulated datasets and real observational data collected during Mini-SiTian test runs.

\subsection{Flare Triggering Simulation Test}
Stellar flares are one of the most common transient sources observed in the Mini-SiTian survey and represent a primary scientific objective during its test runs. During these test runs, upon detecting a flare, the STRIP pipeline triggers a response, initiating follow-up observations with larger telescopes. Therefore, ensuring the consistent and accurate detection of stellar flares by the STRIP pipeline is crucial. However, due to the limited number of flare events observed during testing, a statistical analysis to evaluate the pipeline's performance was not feasible.

To assess the pipeline's capabilities across a range of signal-to-noise ratios (SNR) of host stars and flare intensities, we generated simulated observation images based on real Mini-SiTian data. These simulations closely replicate actual observational conditions (see Figure 3) and feature stars exhibiting flares of varying intensities. By incorporating these simulated images into the standard observational workflow of the STRIP pipeline, we evaluated its flare detection rate and overall efficiency.

As part of our workflow, a dedicated real-bogus classifier (Shi et al., 2025) is used to effectively filter out false detections. Once the classifier removes a significant number of bogus objects, the performance of the test is primarily influenced by the recall rate. Consequently, the focus of this evaluation was on measuring the recall of flare detections, with precision not considered in this particular analysis.

Our simulation involved the creation of two distinct images: one containing flare events and the other without, replicating actual observational conditions. To simulate flare events, we augmented the flux of stars in the second image. These simulated images were then processed through the STRIP pipeline. The pipeline first generates a subtracted image and subsequently searches for variable sources, identifying flare stars. By comparing the number of flare stars detected by the STRIP pipeline to the total number introduced, we assessed the pipeline’s capabilities.

The process for generating simulated observational images is outlined as follows:

1. A precise flux map of stars and background within F02 was obtained by stacking nearly 100 high-quality images. The gain of the images was taken into account to calculate the precise flux at each pixel.

2. A bright star with a high SNR, extracted from the field after background subtraction, was used as a template to simulate other stars.

3. This template star was scaled by varying factors to produce host stars with SNRs of 0.5, 2.5, 10, 33, 100, 300, 1000, and 2000. Additionally, the peak brightness of these stars was increased by 30, 60, 90, 120, 180, 240, 360, and 480 ADU to simulate flare intensities. These stars were then placed in empty regions of the image to generate new flux maps.

4. To account for Poisson noise, we randomly generated noisy observational datasets based on the flux map, incorporating the flare stars. This process was repeated 100 times for each pre- and post-flare image pair. Each pair was processed through the image subtraction and source-finding routines of the STRIP pipeline. The detection rate of flare stars was then determined for fixed host star SNR-flare intensity combinations. This experiment also allowed for the evaluation of the pipeline’s ability to detect supernovae, testing its effectiveness in galaxies with varying surface brightness.

Figure 4 illustrates the detection rates for various host star SNRs and flare intensities in a single observation. The x-axis represents the SNR of the host star where the flare occurred, while the y-axis represents the increase in the peak brightness of the host star in ADU during the flare event. A fitted distribution of these detection rates provides an estimate of the STRIP pipeline’s capability to detect flares, offering valuable insights for future observations.

The figure reveals two regions of low detectability: one at the bottom and one on the right. The bottom region corresponds to flare intensities too faint to be detected in the subtraction image. The right region is saturated because the stars are so bright that their signal exceeds the upper limit of the CMOS detector.

\subsection{Supernova/TDE Observation Trigger Test}
Supernovae, as non-recurring transients with unpredictable locations prior to their explosion, remain poorly understood, particularly in terms of their early physical processes. To address this knowledge gap, the Mini-SiTian project adopts a staring mode, continuously monitoring a designated patch of sky and enabling rapid response to detect supernovae as early as possible. To evaluate whether the STRIP pipeline fulfills these objectives, this section analyzes data from 87 supernova fields observed during the test run, with particular emphasis on several fields where supernovae or TDEs occurred during the observations.

Between 2023-11-06 and 2024-06-27, 87 fields were observed between 40 and 261 times. Among these, 55 events, distributed across 39 fields, are listed on the \href{https://www.wis-tns.org/}{TNS} website. A closer examination revealed that five events were located near the edge of the images and subsequently fell outside the field of view in later observations after the supernova explosion occurred. Of the remaining 50 events, 12 have been confirmed as supernovae on the TNS website, while 38 remain classified as unclassified astronomical transients (ATs). The average observation intervals for these events range from 2.27 to 5.03 days, with an overall average interval of 3.40 days.

A comprehensive comparison of all 50 events within the supernova fields was conducted. Of the 12 confirmed supernovae, 2 were automatically detected by the STRIP pipeline, while 3 out of 38 unclassified astronomical transients (ATs) were identified by the pipeline. Additionally, the program observed a re-brightening event of a tidal disruption event (TDE 2020afhd). A detailed visual inspection revealed three primary reasons for the failure to detect many of these events:

\begin{itemize}
    \item[{$\bullet$}] \textbf{Small Aperture of the Mini-SiTian Telescope: }
    The Mini-SiTian telescope has a relatively small aperture of only 30 cm, significantly smaller than most other time-domain surveys such as ATLAS and ZTF. Combined with an average observation interval of 3.4 days, this resulted in some targets being observable at peak luminosity but becoming too faint for detection after several days of brightness decline. Visual inspection showed that only two supernovae had faint signatures in the science images, while the others were too dim to be detected.
    
    \item[{$\bullet$}] \textbf{Limitations of the STRIP Pipeline's Template Images:}
    The STRIP pipeline generates template images by stacking 20 individual images with good seeing conditions. However, during the early stages of the observation process, when fewer than 20 images are available, the pipeline must rely on single images with relatively poor signal-to-noise ratios (SNR) as templates. This limitation explains why two supernovae, initially visible in the science images, became undetectable after image subtraction. Additionally, early images may not have had optimal seeing conditions. If the science images had better seeing than the template images, the image subtraction process performed suboptimally, leading to an increased number of false detections. However, once the SiTian project reaches full operational capacity, it will conduct sky surveys every 30 minutes, enabling the collection of 20 high-quality images within a few days. This operational enhancement will address the issues encountered during the Mini-SiTian test run.
    
    \item[{$\bullet$}] \textbf{Imaging Issues Hindering Source Subtraction:}
    Some sources could not be effectively subtracted due to imaging issues, such as focusing problems or tracking inaccuracies, which hindered the detection of supernovae.
\end{itemize}

Figure 5 presents a comparison of the images of supernova 2024hrs and TDE 2020afhd, both before and after the image subtraction process within the STRIP pipeline. The corresponding light curve, shown as blue scatter points in the figure, was measured by the STRIP pipeline at the same field. The green dashed lines indicate the times when the supernova and TDE triggered an alarm within the pipeline. For the supernova event SN2024hrs, it is evident that the STRIP pipeline triggered the alarm approximately two days earlier than the earliest report on the TNS. However, it is important to note that the program was not fully operational at the time of the supernova explosion. This early detection was realized only during post-completion testing of the program. 

\begin{figure*}
   \centering
    \includegraphics[width=14cm]{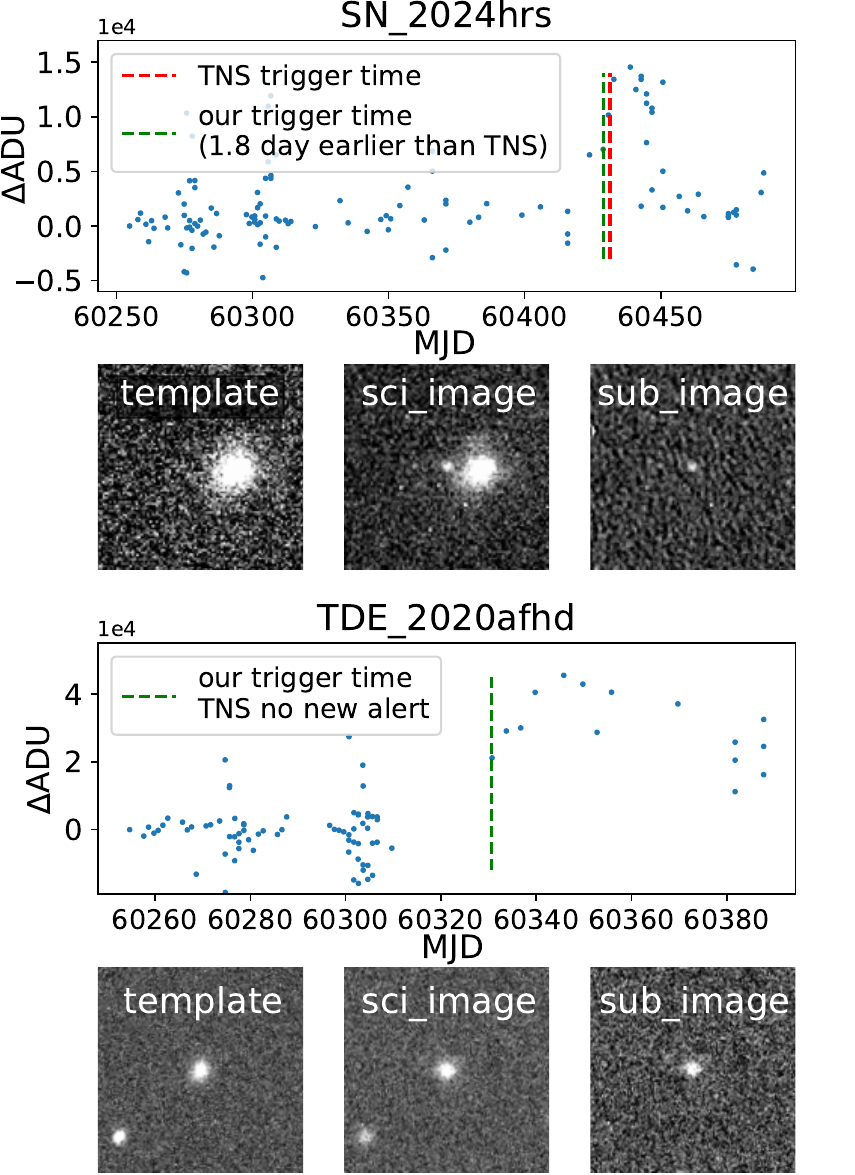}
   \caption{
   Detections and lightcurves of supernova (top two columns) and TDE (bottom two columns) by STRIP. From left to right, it displays the template image, the science image, and the subtraction image. Additionally, the figure shows the light curves from photometry at fixed positions by the STRIP pipeline, in ADU units. The times of the TNS trigger and the time when the transient was detected by the STRIP pipeline are also indicated. It should be noted that the actual output of the STRIP pipeline only includes the light curve after the trigger time, which is presented in multiple formats, including both ADU units and magnitude.
   }
   \label{}
\end{figure*}

\begin{figure*}
   \centering
    \includegraphics[width=13cm]{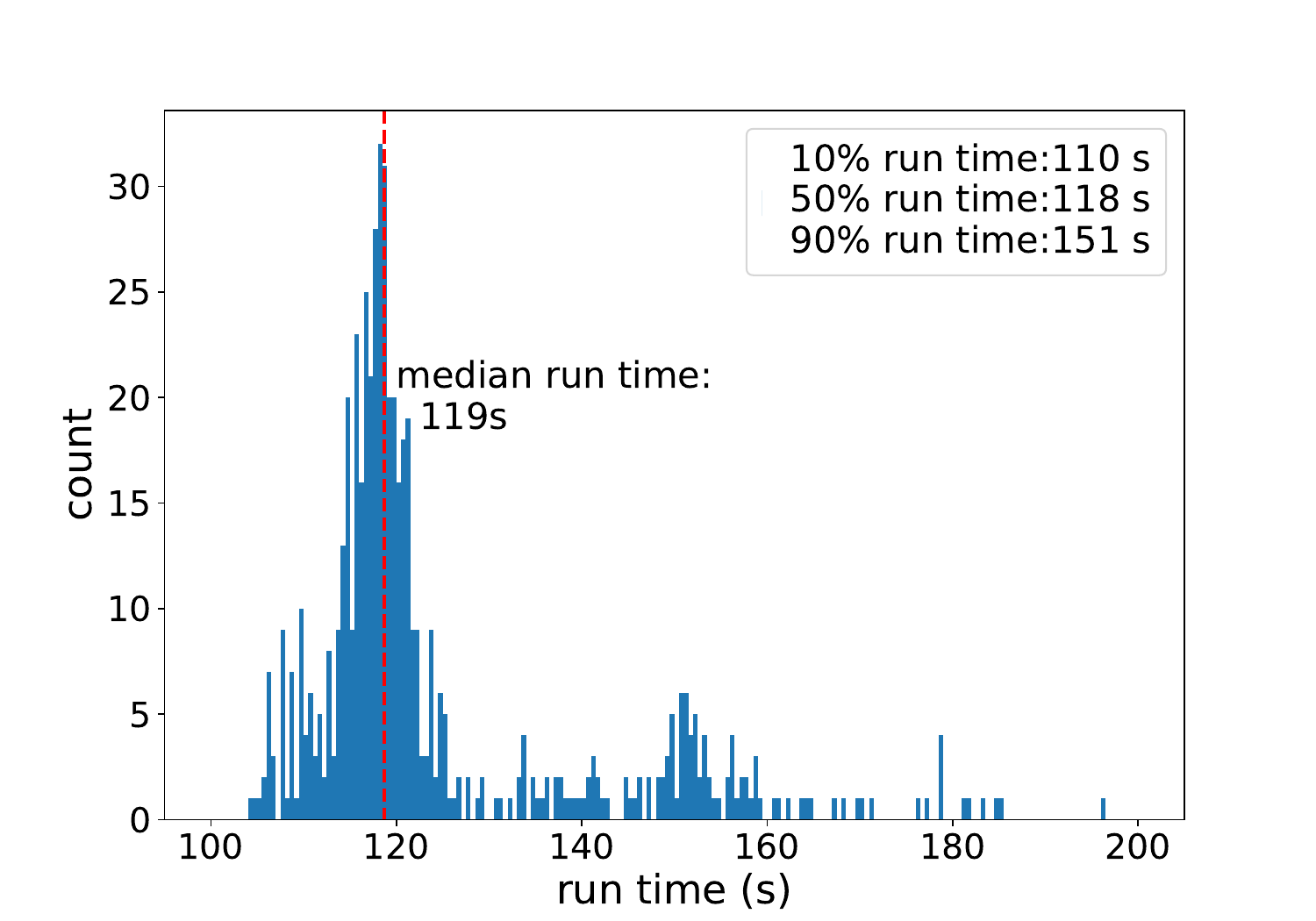}
   \caption{
    Run time distribution of the STRIP pipeline for 582 test images in supernova fields. The processing time of almost all images is within 180s (3 minutes), except for a few (less than 10 images) which are slow in astrometry due to bad weather.
    }
   \label{}
\end{figure*}
\begin{figure*}
   \centering
    \includegraphics[width=14.5cm]{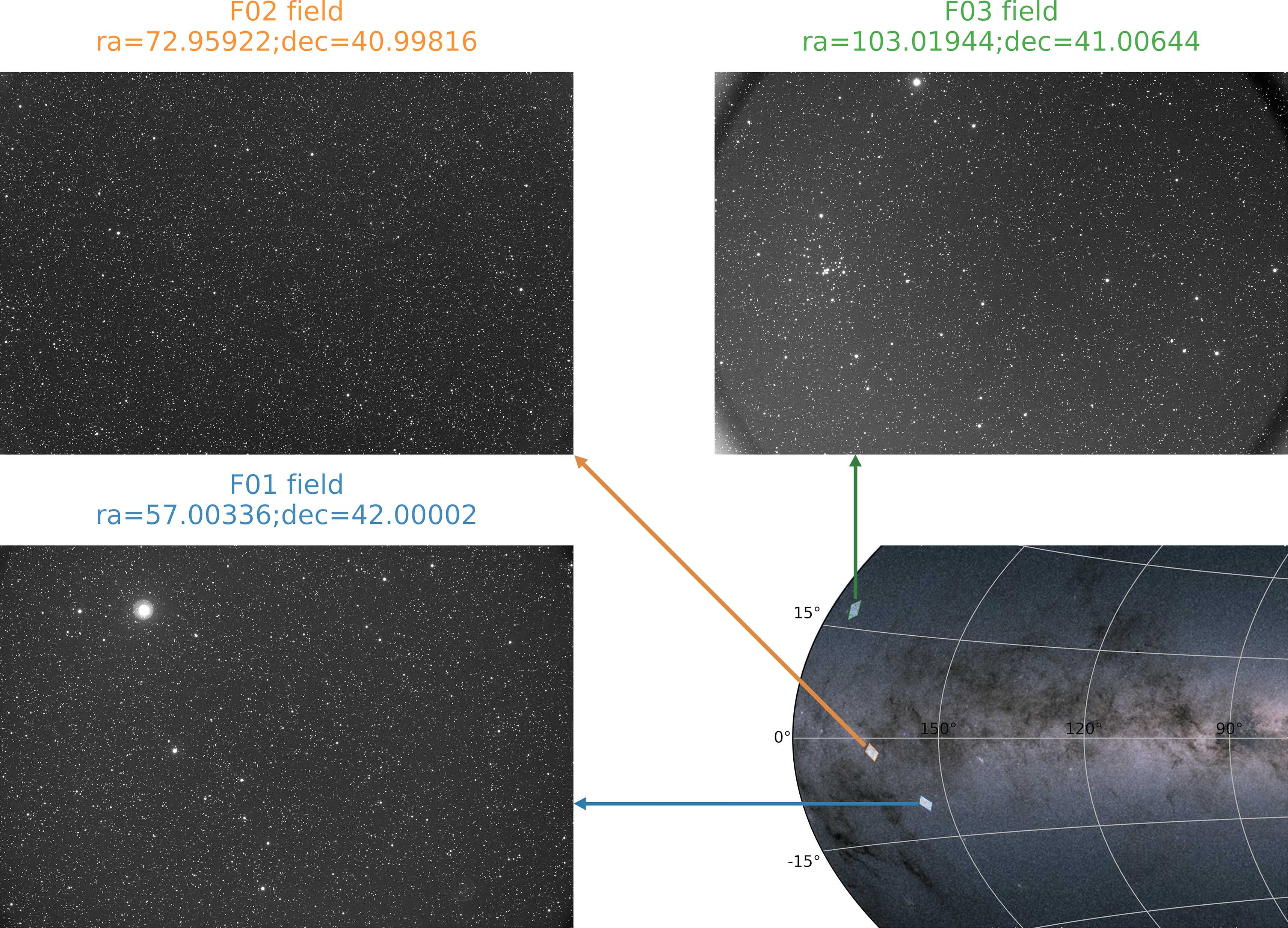}
   \caption{
   Spatial distribution of the three staring mode test field of Mini-SiTian. The lower left (blue), upper left (orange) and upper right (green) are F01, F02 and F03 sky areas respectively. The coordinate system used in the lower right figure is galactic coordinate system, showing the locations of the three fields in the sky.
   }
   \label{}
\end{figure*}
\subsection{Program Execution Efficiency Test}
The testing of supernova fields involved a comprehensive evaluation of the entire STRIP pipeline workflow, simultaneously assessing its operational efficiency. Figure 6 presents the processing times for a total of 582 images captured across 13 supernova fields during the test run. The median processing time was 119 seconds, with 90\% of tasks completed within 152 seconds. Currently, the algorithm does not utilize multi-core parallel computing or GPU acceleration; however, it already meets the requirement of responding within 10 minutes. Looking ahead, with the anticipated implementation of more efficient algorithms, we expect processing speeds to improve significantly, potentially increasing several-fold.

\section{Data Products During the Test run}
During the test run, in addition to observing supernovae to validate our ability to detect long-timescale transient sources, we selected three sky fields with midnight culmination near the zenith to assess the performance of prolonged continuous staring observations. These fields, sequenced according to their observation schedules, were designated as field-01 (F01), field-02 (F02), and field-03 (F03). Observations in these fields were conducted using both the Mini-SiTian 2 telescope (SDSS-g band) and the Mini-SiTian 3 telescope (SDSS-r band). The central coordinates and field-of-view ranges for these three fields are illustrated in Figure 7. Specifically, F01 was observed 1,956 times by the Mini-SiTian 2 telescope and 1,469 times by the Mini-SiTian 3 telescope. F02 was observed 3,636 times by the Mini-SiTian 2 telescope and 3,554 times by the Mini-SiTian 3 telescope, while F03 was observed 412 times by the Mini-SiTian 2 telescope and 491 times by the Mini-SiTian 3 telescope.

To evaluate the variable source detection capability of the STRIP pipeline, all dual-band images from F02 were thoroughly scrutinized, with every source triggering an alert logged. In the absence of an automated light curve classification system, images containing more than 100 alerts over the course of 3,500 observations were flagged as potential variable source candidates, resulting in a total of 421 objects. These candidates were subsequently manually classified through visual inspection. Ultimately, the following variable sources were identified within F02: 1 flare event, 49 eclipsing binaries, and 78 pulsating variables. A comparison with existing variable star catalogs, such as those from ATLAS (\citealt{2018PASP..130f4505T}) and \href{https://simbad.u-strasbg.fr/simbad/}{SIMBAD} Database, revealed that 9 pulsating variables were not present in these catalogs and were thus classified as new variable star candidates. 

Two distinct datasets were compiled to facilitate the subsequent training of machine learning algorithms that will be integrated into the STRIP pipeline:

\begin{figure*}
   \centering
    \includegraphics[width=14cm]{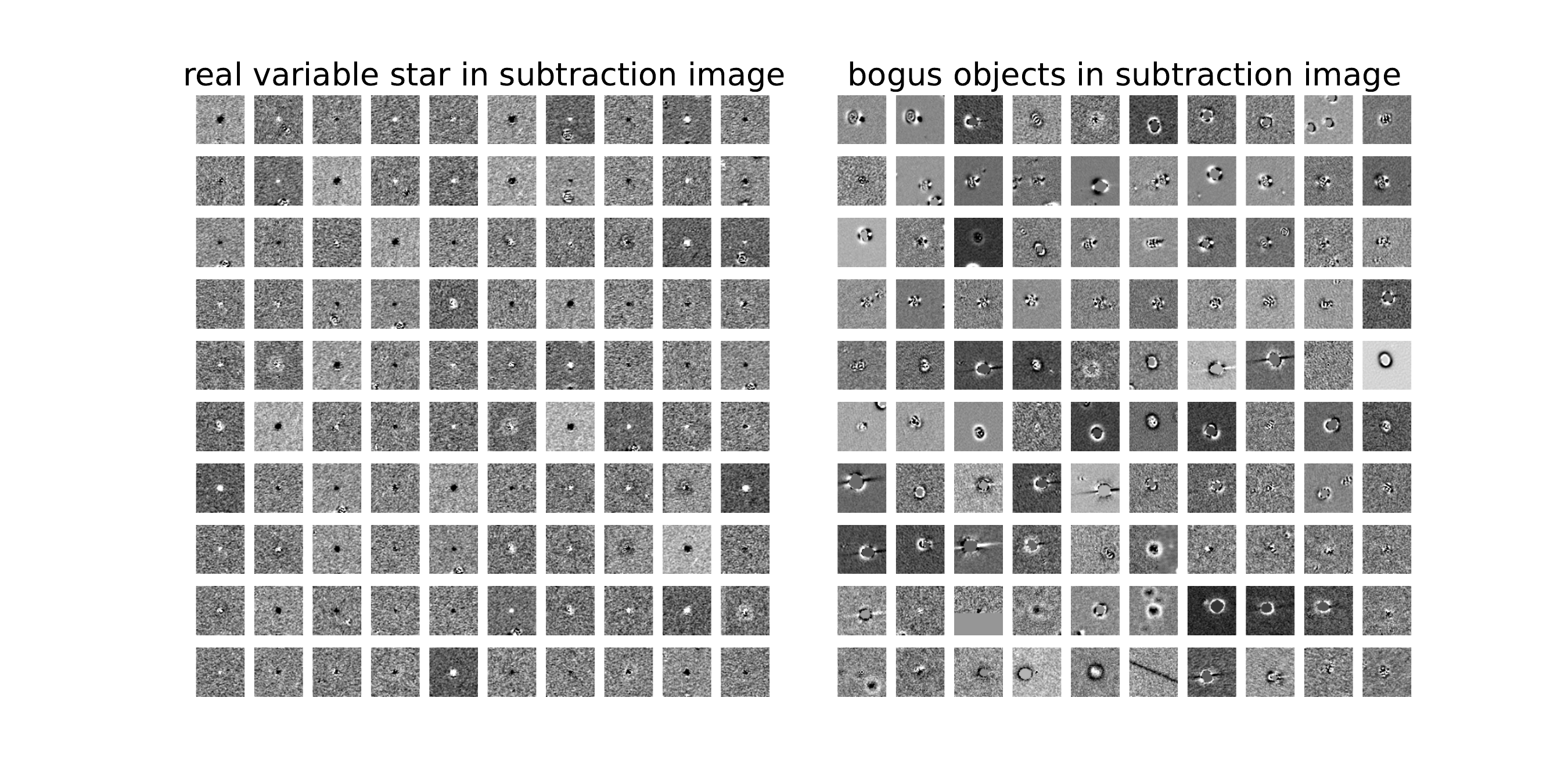}
   \caption{
   Random selections of 100 examples of real variables (left panel) and bogus ones (right panel) respectively.
   }
   \label{}
\end{figure*}
\begin{figure*}
   \centering
    \includegraphics[width=14cm]{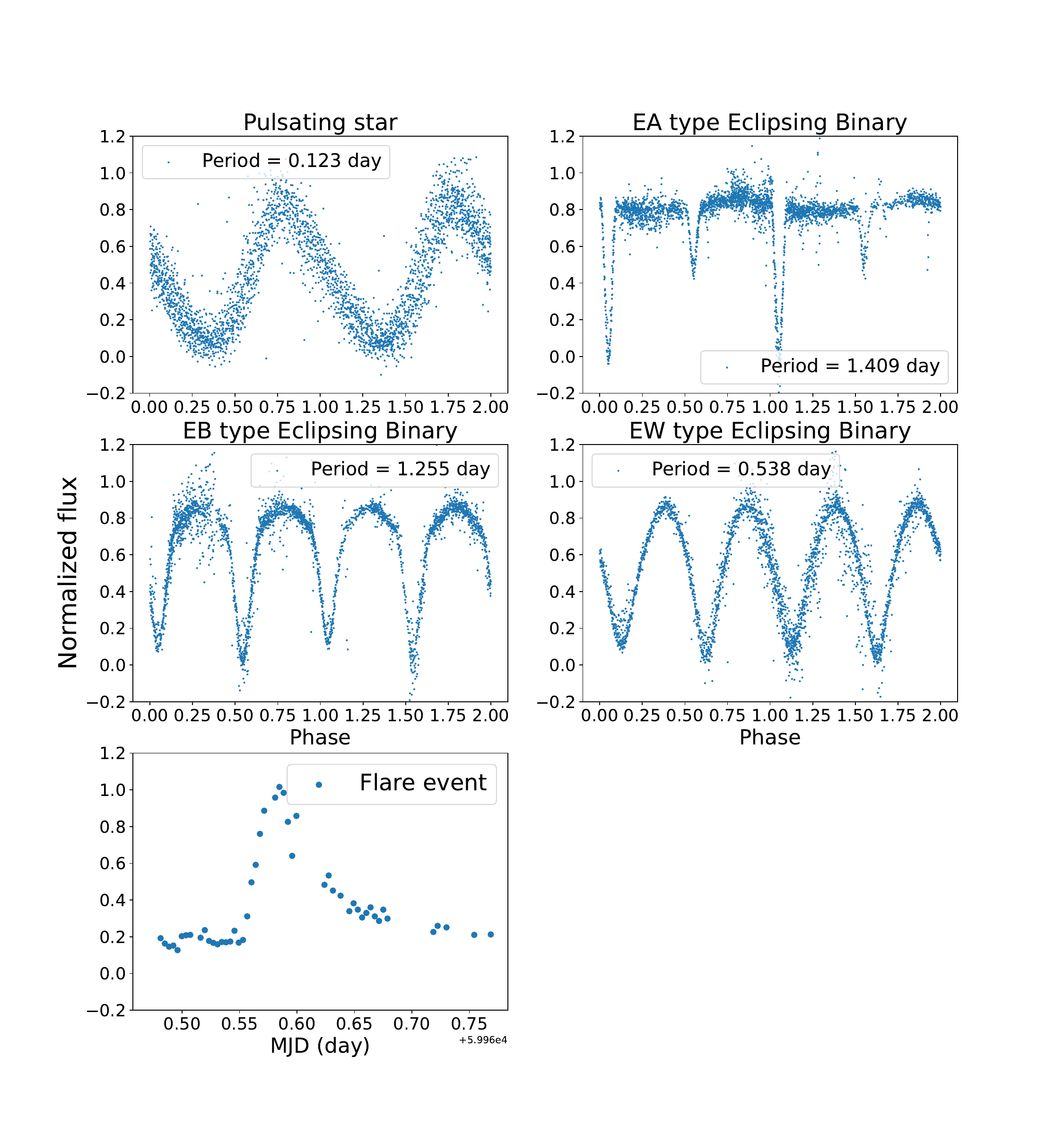}
   \caption{
   Examples of the light curves of the 5 typical variable sources seen in the F02. All light curves have been normalized to a range of 0 to 1 using their respective 5th and 95th percentiles as the y-axis coordinates. The x-axis for the four upper panels, which display period-folded light curves, is given in units of phase. For the stellar flare light curve shown in the lower panel, the x-axis is represented by Modified Julian Date (MJD). The light curves of all manually classified variable sources have been labeled and added to the training dataset of the light curve classifier algorithm.
   }
   \label{}
\end{figure*}

1) Training Dataset for Real-Bogus Classifiers: A dataset was created by cropping 64$\times$64 pixel sub-images centered on the positions of variable sources exhibiting significant variations (real sources) and at locations incorrectly identified as variable sources (bogus sources) in subtracted images. The training dataset includes approximately 4,000 positive cases and 215,000 negative cases. A sample of 100 examples from both categories is shown in Figure 8.

2) Light Curve Dataset for Classifier Training: Targets that triggered more than 100 alerts were classified as variable stars, with corresponding labels assigned based on visual inspection. A total of 127 light curves with obvious periodic signals were labeled as either "pulsating stars" or "eclipsing binaries," while 293 light curves exhibiting pure noise were labeled as "non-variable sources." These labeled light curves now form a dataset for training an automatic variable source classifier. A selection of five example light curves is presented in Figure 9.

Asteroids passing through the observed fields may trigger single responses without generating multiple alerts. By plotting all response triggers, using different colors to represent different observation times within a single figure, we can visualize a series of colorful lines that facilitate asteroid detection, independent of reliance on an asteroid database. Utilizing this method, we successfully identified 14 asteroids in field F02, which represents more than half of the 26 asteroids, brighter than magnitude 20, predicted to be present in this region according to the orbital parameters provided in the Mini-SiTian asteroid analysis paper (Liu et al., 2025).
Upon further inspection, all 11 asteroids brighter than magnitude 19 were detected using this approach. This outcome demonstrates the effectiveness of rapidly identifying asteroids directly from the variable source candidate catalogs generated by the STRIP pipeline. When compared with the known asteroid database, all detected asteroids were found to be listed in the Minor Planet Center (MPC) database.

\section{Limitation and Improvement}
Currently, the Mini-SiTian array operates with a single node containing three telescopes for test observations, generating three 6k $\times$ 
9k images every five minutes. Under the current CPU configuration, the Hotpants software performs satisfactorily, completing data processing tasks within the required timeframes. However, when scaling to the full SiTian array, which will consist of over sixty telescopes, operational challenges are expected. Each telescope unit will house four 9k $\times$ 
9k CMOS detectors, collectively producing data at approximately two-minute intervals. At this scale, the existing Hotpants implementation may face processing bottlenecks due to limited computational resource. Potential solutions to address these issues include optimizing Hotpants' runtime efficiency or adopting GPU-accelerated computational algorithms, such as the SFFT algorithm, which can leverage the rapidly advancing capabilities of parallel computing architectures.

In parallel, the STRIP pipeline successfully generates numerous transient candidate light curves in real-time during its operation. Further processing with a machine learning classifier or large language model, such as Falco (Zuo et al., in prep), will utilize these multi-band information to classify transients based on their light curves. Once integrated into the STRIP pipeline, this system will enhance the pipeline’s capabilities, enabling real-time transient alerts and automated source classification.
\section{Conclusion and Prospects}
To benchmark the scientific objectives of the SiTian project using the Mini-SiTian telescopes, we developed the STRIP pipeline. This pipeline is specifically designed to enable near real-time detection of short-timescale transient events, such as stellar flares, within ten minutes of observation, while also handling long-duration transients like supernovae and gravitational wave electromagnetic counterparts. Our testing has confirmed the STRIP pipeline’s excellent performance in the staring mode observation. Using data from supernova and TDE fields, we validated its efficiency in detecting these transients. Additionally, data from the F02 field in the test run provided insights into the reliability of the STRIP pipeline, resulting in two datasets that will be used to train advanced machine learning modules for future integration into the pipeline.

Building on the data accumulated from the initial test run, the STRIP pipeline is now prepared to tackle further challenges in the upcoming second round of Mini-SiTian observations, scheduled for September 2024. We anticipate continuous iterations and optimizations, with the goal of significantly advancing the scientific outcomes expected from the Mini-SiTian project.

\subsection*{acknowledgements}
The SiTian project is a next-generation, large-scale time-domain survey designed to build an array of over 60 optical telescopes, primarily located at observatory sites in China. This array will enable single-exposure observations of the entire northern hemisphere night sky with a cadence of only 30-minute, capturing true color (gri) time-series data down to about 21 mag. This project is proposed and led by the National Astronomical Observatories, Chinese Academy of Sciences (NAOC). As the pathfinder for the SiTian project, the Mini-SiTian project utilizes an array of three 30 cm telescopes to simulate a single node of the full SiTian array. The Mini-SiTian has begun its survey since November 2022. The SiTian and Mini-SiTian have been supported from the Strategic Pioneer Program of the Astronomy Large-Scale Scientific Facility, Chinese Academy of Sciences and the Science and Education Integration Funding of University of Chinese Academy of Sciences.

Y.H. acknowledges the supports from the National Key Basic R\&D Program of China via 2023YFA1608303, the Strategic Priority Research Program of the Chinese Academy of Sciences (XDB0550103) and the National Science Foundation of China (NSFC Grant Nos. 12422303 and 12261141690). K.X. acknowledges the supports from the NSFC grant No. 12403024, the Postdoctoral Fellowship Program of CPSF under Grant Number GZB20240731, the Young Data Scientist Project of the National Astronomical Data Center, and the China Post-doctoral Science Foundation No. 2023M743447. Z.F. acknowledges the supports from the Strategic Priority Research Program of the Chinese Academy of Sciences，Grant No. XDB0550000. J.F.L. acknowledges support the NSFC through grant Nos. of 11988101 and 11933004, and support from the New Cornerstone Science Foundation through the New Cornerstone Investigator Program and the XPLORER PRIZE.

\bibliographystyle{raa}
\bibliography{RAA-2024-0355.R1}

\label{lastpage}

\end{document}